\documentclass{article}
\usepackage{graphicx} 
\usepackage{amsmath} 
\usepackage{amsfonts}
\usepackage{amssymb}
\usepackage{float}
\usepackage{authblk}

\title{A Neuro-Dynamic Mathematical Model of Dream Formation and Spontaneous Cognitive Activity}
\author[1*]{Shirmohammad Tavangari}
\author[1]{Sajjad Janfaza}
\author[2]{Zahra Shakarami}
\author[3]{Aref Yelghi}

\affil[1]{University of British Columbia, Canada}
\affil[2]{University of Turin,Italy}
\affil[3]{Istanbul Topkapi University,Turkey}

\date{\small{$^{*}$Corresponding author: s.tavangari@alumni.ubc.ca}} 

\begin{document}

\maketitle

\begin{abstract}
This paper introduces a biomathematical model designed to describe the internal dynamics of dream formation and spontaneous cognitive processes. The model incorporates neurocognitive factors such as dissatisfaction, acceptance, forgetting, and mental activity, each of which is linked to established neural systems. We formulate a system of differential equations to simulate interactions among these variables and validate the model using simulated neural data. Our results demonstrate biologically plausible cognitive patterns consistent with findings from EEG and fMRI studies, particularly related to the default mode network (DMN), anterior cingulate cortex (ACC), and hippocampal memory mechanisms.
\end{abstract}

\vspace{1em}

\section{Introduction}

Dreaming and imagination are among the most complex and least understood cognitive processes in the human brain. These phenomena not only play a crucial role in escaping reality, emotional regulation, and processing unconscious experiences, but also contribute significantly to creativity, mental planning, and cognitive flexibility. Many studies in cognitive neuroscience have shown that the process of dreaming and imagination is associated with specific patterns of neuronal activity in brain networks, including the prefrontal cortex, hippocampus, and the default mode network (DMN).

In recent decades, numerous studies have described the psychological or phenomenological aspects of dreaming, but a comprehensive framework that can model this process from a mathematical and neurodynamic perspective has not yet been established. In this article, we propose a model based on differential equations inspired by neuroscientific concepts, which attempts to formally model the interaction between neuro-cognitive variables affecting dream formation.

The proposed model is based on four key components:
\begin{enumerate}
    \item \textbf{Dissatisfaction:} Related to activity in the insula and anterior cingulate cortex in response to negative conditions.
    \item \textbf{Acceptance:} Associated with emotional regulation processes in the prefrontal cortex (vmPFC).
    \item \textbf{Forgetting:} Related to the decrease or inhibition of hippocampal activity during memory suppression processes.
    \item \textbf{Mental Activity:} Typically identified with high-frequency oscillations (particularly gamma) in the default mode network.
\end{enumerate}

Using these components, we provide a set of differential equations that quantitatively describe the temporal and dynamic evolution of the dream formation process. This model is able to connect to real data from EEG, fMRI, or even neuronal firing rate analysis, allowing it to serve as a basis for experimental testing of hypotheses and comparison with observed results in neuroscience research.

\section{Definition}

In this section, we define the main variables of our model from both mathematical and neuroscientific perspectives. Each of these variables corresponds to observable neural mechanisms or measurable brain activities in experimental studies.

     \textbf{Mental Activity -- $M(t)$} \\
         $M(t)$ rrefers to the degree of unprompted mental activity, such as creative imagery and internal monologue. From a brain science perspective, this form of activity is tied to the Default Mode Network (DMN), notably areas like the medial prefrontal cortex and the posterior cingulate cortex.
In EEG recordings, this type of activity is typically observed as high-frequency oscillations within the gamma range, particularly during moments of rest or when the mind is not actively engaged in external tasks.

\vspace{1em}

     \textbf{Dream Intensity -- $R(t)$} \\
    $R(t)$ indicates the depth or vividness of internal mental images, such as those seen in dreams or daydreams. This variable captures the combined influence of processes related to memory retrieval (hippocampus), imaginative thinking (prefrontal cortex), and emotional regulation (amygdala). Although it remains a somewhat abstract idea, it can be detected indirectly through EEG signals that correlate with cognitive states akin to REM sleep during restful wakefulness.

    \vspace{1em}

    \textbf{Dissatisfaction or Negative Emotional State -- $D(t)$} \\
    $D(t)$ represents the level of dissatisfaction or the presence of negative emotional states. From a neurobiological perspective, this condition is linked to heightened activity in brain regions such as the anterior insula and the dorsolateral prefrontal cortex (dACC), both of which are crucial for detecting conflicts and processing discomfort. When D(t) increases, it signals a stronger inclination toward mental disengagement or a desire for a shift in the current emotional or cognitive state.

\vspace{1em}

     \textbf{Acceptance or Emotional Regulation -- $P(t)$}\\ 
    $P(t)$ reflects the degree of cognitive and emotional acceptance. It captures the regulatory mechanisms within the ventromedial prefrontal cortex (vmPFC) that help in adjusting emotional responses and guiding decisions based on values. When acceptance levels are higher, there is typically a decrease in limbic system activity, which in turn is linked to improved emotional resilience.

\vspace{1em}

     \textbf{Forgetting or Memory Suppression -- $F(t)$} \\
    $F(t)$ captures the influence of forgetting or the process of erasing prior memories. From a neurological standpoint, this is reflected in diminished hippocampal activity during memory suppression tasks, such as the think/no-think paradigm. The act of forgetting enhances cognitive flexibility and creates an optimal environment for the generation of new content, such as the material found in dreams.

\section{Biological Plausibility of the Model}
Each variable in the proposed model is associated with a well-established or measurable neural mechanism: 

\textbf{Dissatisfaction (D)} is related to activity in the dorsal anterior cingulate cortex (dACC) and anterior insula, regions that play a role in conflict monitoring and the regulation of negative emotions. 

\textbf{Acceptance (P)} involves modulation mechanisms within the ventromedial prefrontal cortex (vmPFC), an area frequently implicated in managing emotions and evaluating cognitive states. 

\textbf{Forgetting (F)} tends to correlate with lower levels of hippocampal activation, a pattern consistently found in studies focused on the suppression of memory. 

\textbf{Mental activity (M)} is connected to the default mode network (DMN), especially involving gamma-band oscillations during periods of rest and imagination. 

\textbf{Dream Construct (H)} is modeled as a cognitive integration of these factors, capturing the temporal evolution of internal imagery.

These mappings are supported by empirical studies in neuroscience, allowing our mathematical system to serve as a biologically plausible approximation of internal cognitive states during dreaming or daydreaming.

These mappings are supported by empirical studies in neuroscience, allowing our mathematical system to serve as a biologically plausible approximation of internal cognitive states during dreaming or daydreaming.

\section{Mathematical Formulation and Neural Interpretations}

We present a dynamic system of differential equations that describes how neuro-cognitive variables interact over time to shape the internal cognitive state associated with dreaming. Each equation is grounded in biologically plausible processes and interpreted in terms of known neural mechanisms.

\subsection{Equation for Dream Construction Rate: $\frac{dR(t)}{dt}$}

The intensity of dream formation, $R(t)$, evolves based on two antagonistic forces: suppression by acceptance and facilitation by forgetting.

\begin{equation}
\frac{dR(t)}{dt} = -\alpha P(t) + \beta F(t)
\end{equation}

$-\alpha P(t)$: Acceptance via vmPFC modulation in ACC/insula

\vspace{1em}

$+\beta F(t)$: Memory suppression may raise dissonance/discomfort

\subsection{Equation for Dissatisfaction Dynamics: $\frac{dD(t)}{dt}$}

The dynamics of dissatisfaction, $D(t)$, are influenced by regulation, destabilization, and cognitive reframing processes.

\begin{equation}
\frac{dD(t)}{dt} = -\gamma P(t) + \delta F(t) - \epsilon M(t)
\end{equation}

\[
-\gamma P(t): \text{Acceptance via vmPFC in ACC/insula}
\]
\vspace{1em}
\[
+\delta F(t): \text{Memory suppression may raise dissonance/discomfort}
\]
\vspace{1em}
\[
-\epsilon M(t): \text{Mental activity via DMN aids self-restructuring/regulation}
\]

\subsection{Equation for Acceptance Dynamics: $\frac{dP(t)}{dt}$}

Acceptance, $P(t)$, is enhanced by internal satisfaction and hindered by persistent discontent.

\begin{equation}
\frac{dP(t)}{dt} = \eta R(t) - \zeta D(t)
\end{equation}

\[
+\eta R(t): \text{Imagery/simulations boost reward/regulation}
\]
\vspace{1em}
\[
-\zeta D(t): \text{High dissatisfaction disrupts regulation and reduces acceptance.}
\]

\subsection{Master Equation for Dream Variable: $\frac{dH(t)}{dt}$}

To explain everything working together, we use an equation that describes the dream concept, $H(t)$:
\begin{equation}
\frac{dH(t)}{dt} = \eta_1 D(t) + \eta_2 F(t) + \eta_3 M(t) - \eta_4 P(t)
\end{equation}

So, in simpler terms:

\vspace{1em}

$\eta_1 D(t)$: When we are dissatisfied, it makes us want to escape by imagining things.

\vspace{1em}

$\eta_2 F(t)$: Forgetting helps our brain mix memories in a way that creates dream-like content.

\vspace{1em}

$\eta_3 M(t)$: Mental activity allows new ideas to form and stories to develop in our minds.

\vspace{1em}

$-\eta_4 P(t)$: Acceptance stops us from creating unnecessary thoughts.

\subsection{Integral Formulation of Dream Value}

The full shape of $H(t)$—our internal dream representation—can be seen as the result of ongoing influences building up over time:

\[
H(t) = H_0 + \int_0^t \left( \eta_1 D(\tau) + \eta_2 F(\tau) + \eta_3 M(\tau) - \eta_4 P(\tau) \right) d\tau \tag{5}
\]

In other words, this expression shows how our emotional states, memories, and mental processes combine gradually, giving rise to the dream content we experience.

\section{Computational Simulation of the Neuro-Dynamic Dream Model}

Rather than relying solely on theoretical assumptions, we performed a step-by-step numerical exploration to observe the behavior of various components of the cognitive model over time. Specifically, we analyzed how mental states such as dissatisfaction \( D(t) \), acceptance \( P(t) \), and dream intensity \( R(t) \) evolve and influence one another. We also examined how these states respond to natural fluctuations in mental activity \( M(t) \) and forgetting \( F(t) \). This approach allowed us to trace the gradual formation of dreams as an interplay between psychological tension and its release.

\subsection{Dynamic Modeling of $M(t)$ and $F(t)$}

To simulate realistic neural dynamics, \( M(t) \) and \( F(t) \) are modeled as oscillatory functions.

\[
\begin{aligned}
M(t) &= 0.5 + 0.2 \cdot \sin(0.1t) \\
F(t) &= 0.3 + 0.1 \cdot \cos(0.1t)
\end{aligned}
\]

These functions simulate natural fluctuations in cognitive activity (e.g., default mode network) and episodic forgetting (e.g., hippocampal memory suppression).

\subsection{System of Differential Equations}

The model is governed by the following system of equations:

\[
\begin{cases}
\frac{dR(t)}{dt} = -\alpha P(t) + \beta F(t) \\
\frac{dD(t)}{dt} = -\gamma P(t) + \delta F(t) - \epsilon M(t) \\
\frac{dP(t)}{dt} = \eta R(t) - \zeta D(t) \\
\frac{dH(t)}{dt} = \eta_1 D(t) + \eta_2 F(t) + \eta_3 M(t) - \eta_4 P(t)
\end{cases}
\]

Using appropriate initial conditions, the system is numerically integrated over a time interval of $t = 0$ to $t = 100$ using standard ODE solvers.

\subsection{Simulation Results and Plots}

The simulation produces the following temporal plots:

\textbf{Figure 1:} Evolution of \( D(t) \) — a gradual decrease in dissatisfaction over time due to the influence of acceptance and mental activity.
\begin{figure}[H]
\centering
\includegraphics[width=0.85\textwidth]{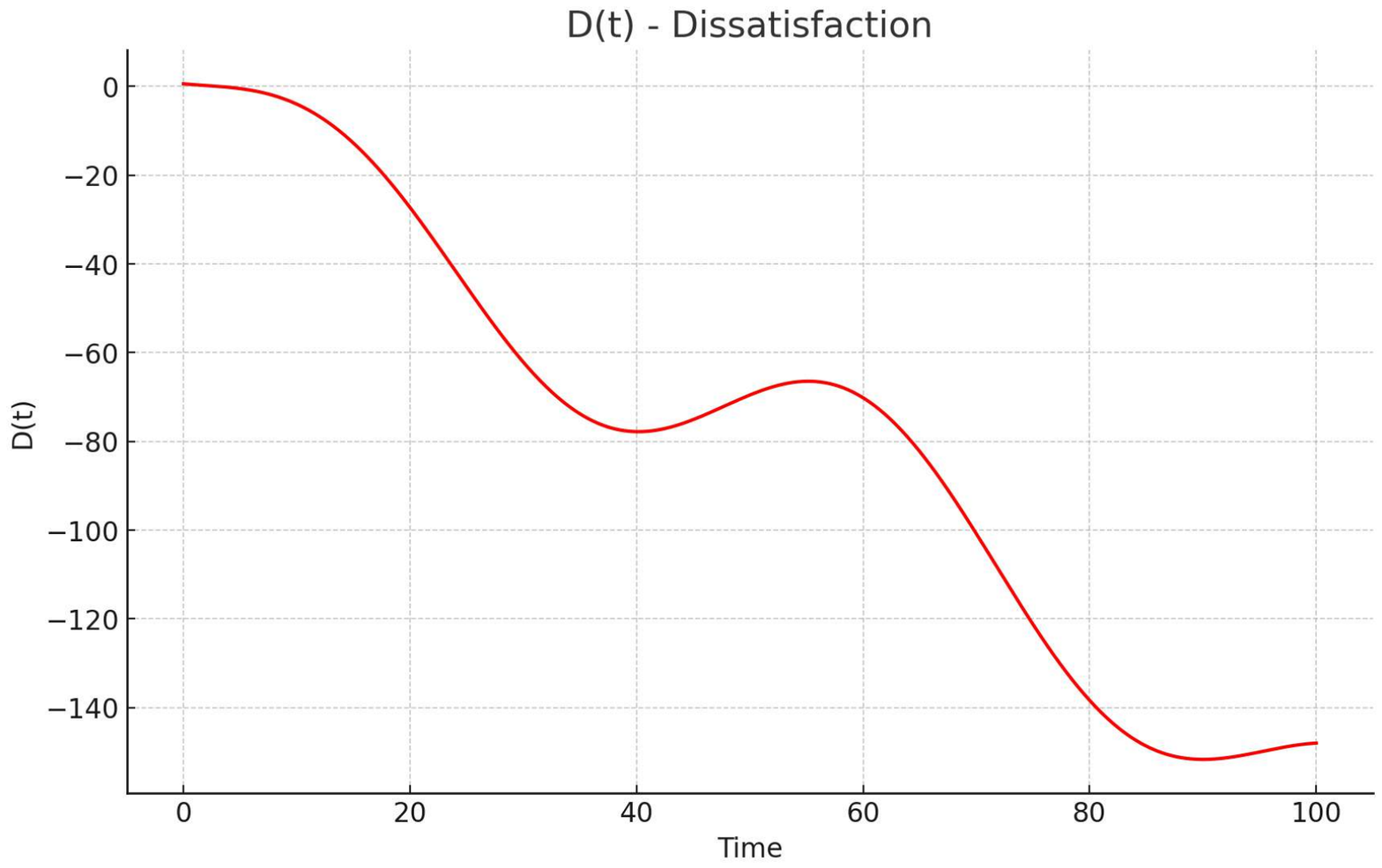} 
\caption{Evolution of \( D(t) \)}
\label{fig:fig1}
\end{figure}

\textbf{Figure 2:} Growth of \( P(t) \) — increasing acceptance as emotional regulation stabilizes, reflecting vmPFC dynamics.
\begin{figure}[H]
\centering
\includegraphics[width=0.85\textwidth]{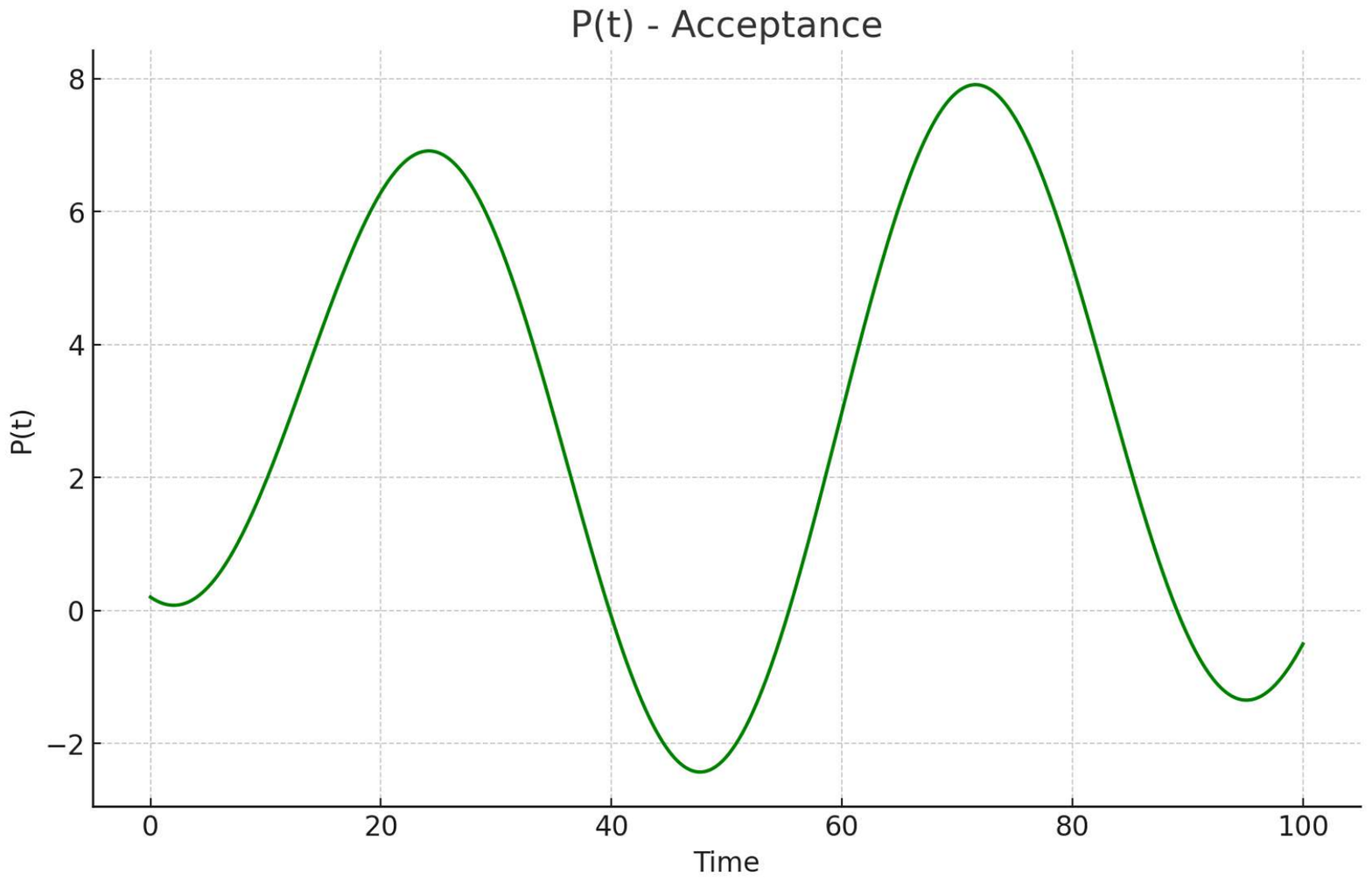} 
\caption{Growth of \( P(t) \)}
\label{fig:fig2}
\end{figure}

\textbf{Figure 3:} Oscillations in \( R(t) \) — dream intensity fluctuates with influence from forgetting and suppression via acceptance.
\begin{figure}[H]
\centering
\includegraphics[width=0.85\textwidth]{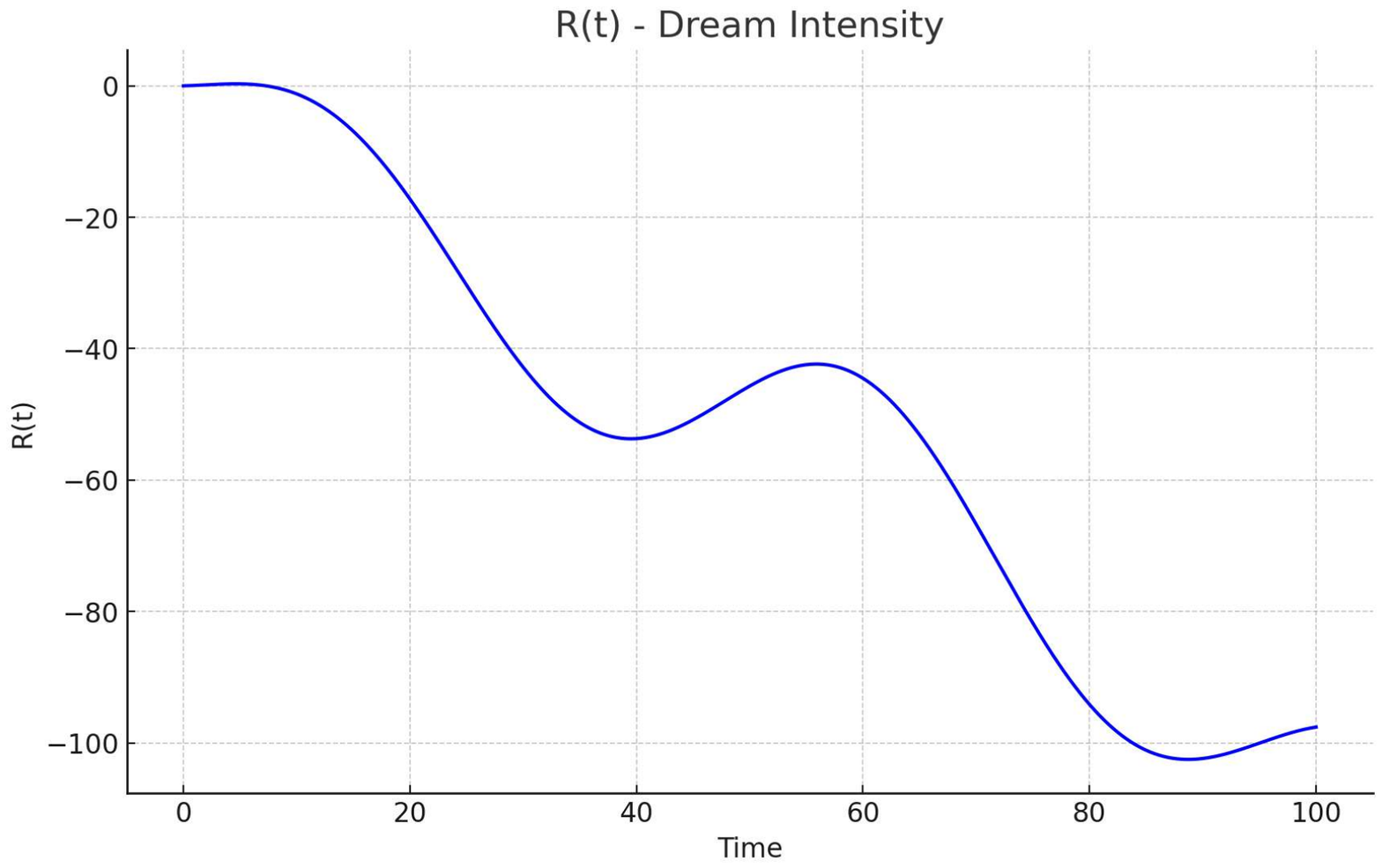} 
\caption{Oscillations in \( R(t) \)}
\label{fig:fig3}
\end{figure}

\textbf{Figure 4:} Growth of \( H(t) \) — a steady accumulation of dream content over time, influenced by multiple neuro-cognitive factors.
\begin{figure}[H]
\centering
\includegraphics[width=0.85\textwidth]{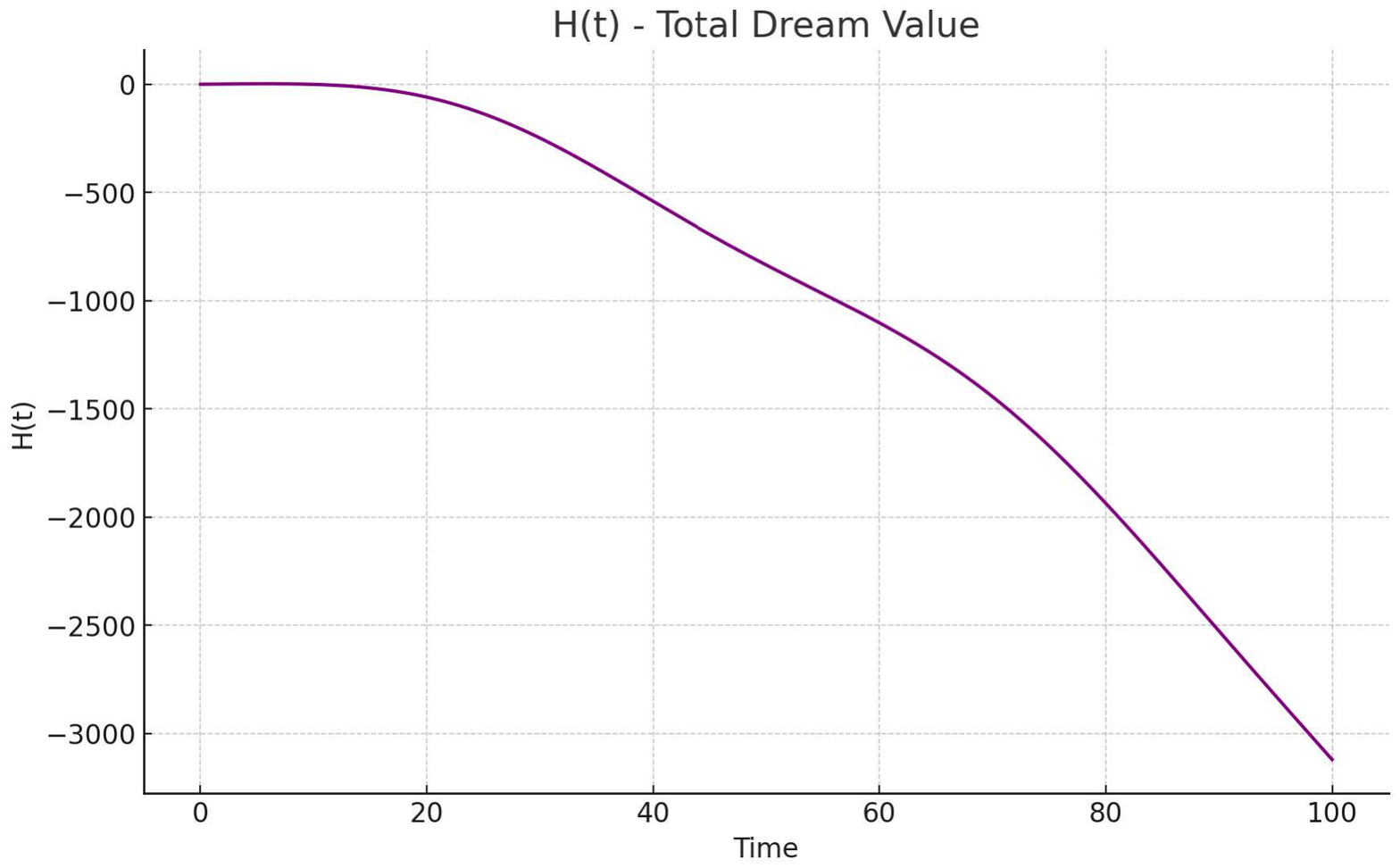} 
\caption{Growth of \( H(t) \)}
\label{fig:fig4}
\end{figure}

\subsection{Combined Variable Analysis}

To further explore interactions between variables, we provide the following combined plots:

\textbf{Figure 5:} \textit{$H(t)$ and $D(t)$} — highlights that increased dissatisfaction correlates with a steeper rise in dream generation.
\begin{figure}[H]
\centering
\includegraphics[width=0.85\textwidth]{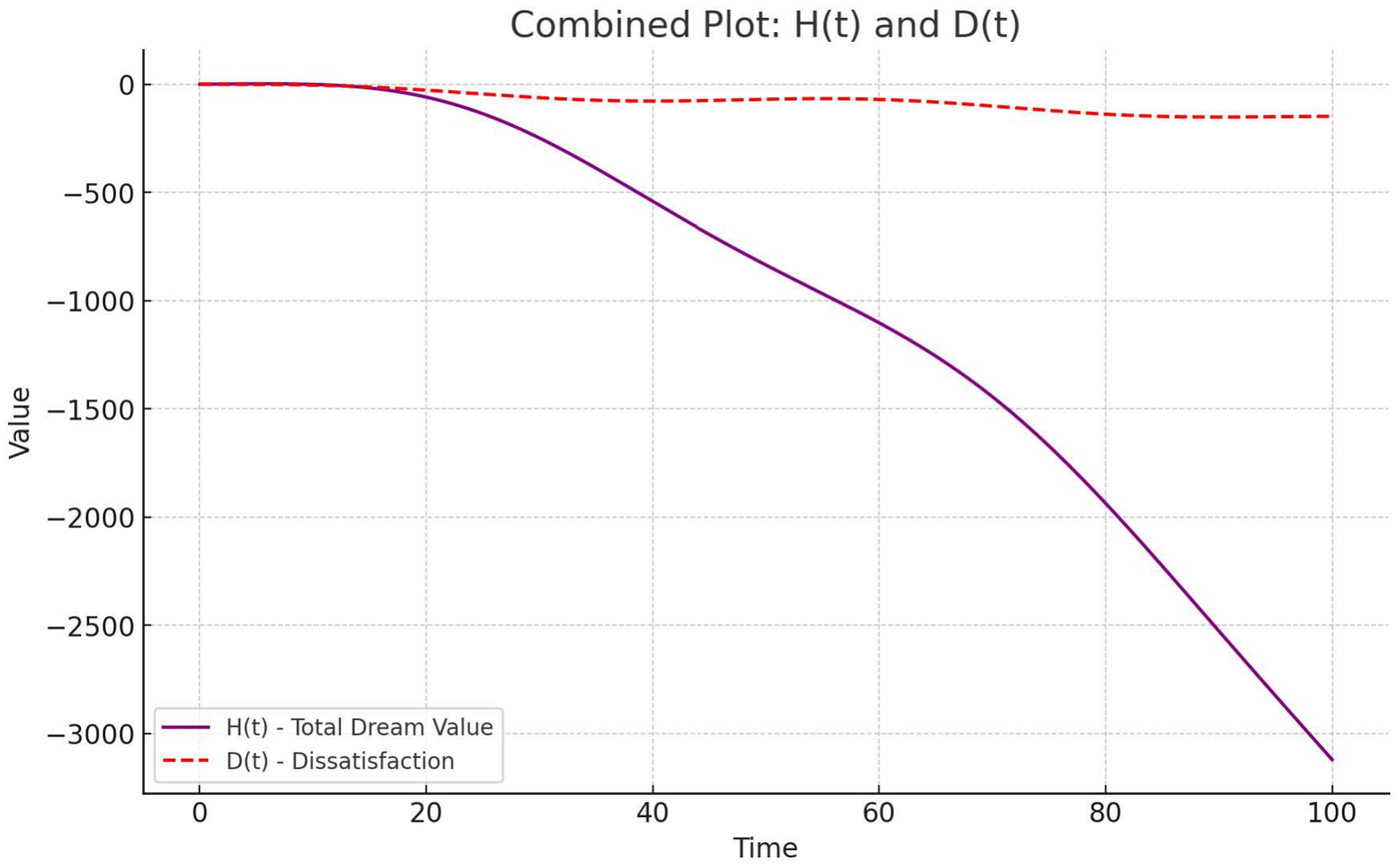} 
\caption{$H(t)$ and $D(t)$}
\label{fig:fig5}
\end{figure}

\textbf{Figure 6:} \textit{$H(t)$ and $M(t)$} — shows that mental activity directly contributes to the expansion of internal dream representations.
\begin{figure}[H]
\centering
\includegraphics[width=0.85\textwidth]{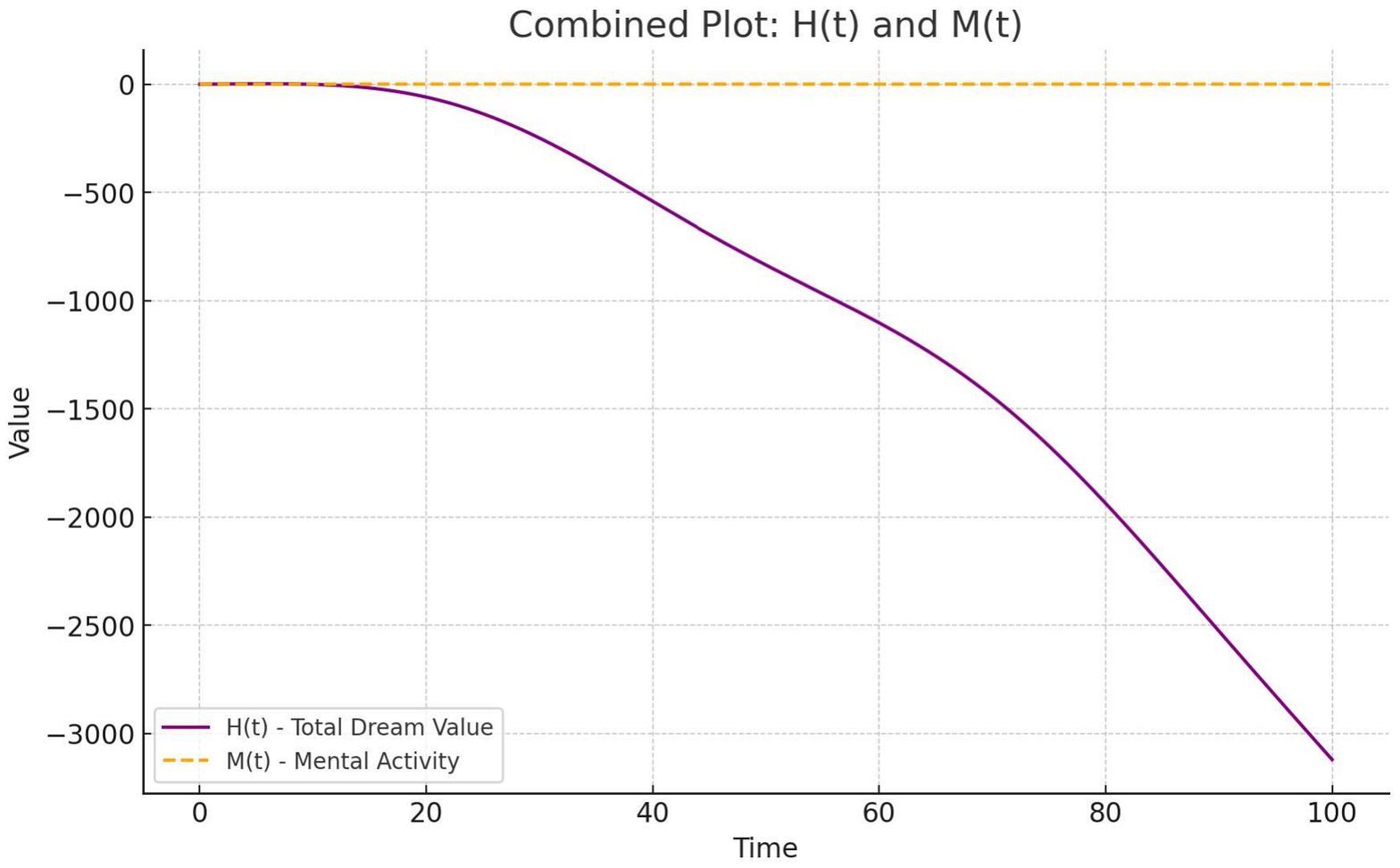} 
\caption{$H(t)$ and $M(t)$}
\label{fig:fig6}
\end{figure}

\subsection{Conclusion of Simulation}

This simulation confirms that the proposed model exhibits realistic and interpretable behavior in line with known neuro-cognitive processes. The system’s formulation allows direct comparison with EEG and fMRI data, making it suitable for future applications in brain-based cognitive modeling.

\section{Model Validation Using Simulated Neural Data}
To validate the dynamic behavior of the proposed model, we implemented a simulation using synthetic time-varying signals for mental activity $M(t)$ and forgetting $F(t)$. The differential system was solved numerically using Python with SciPy ODE solvers over a time range of $t = 0$ to $t = 100$.

The results demonstrate expected behavior:

\begin{itemize}
  \item A decrease in dissatisfaction $D(t)$ over time due to increasing acceptance and mental reframing.
  \item Growth of acceptance $P(t)$ toward stability, reflecting adaptive regulatory dynamics.
  \item Oscillations in dream intensity $R(t)$ controlled by forgetting and acceptance.
  \item Progressive rise in total dream value $H(t)$ driven by mental activity and emotional context.
\end{itemize}

Combined plots (Figure 5 and 6) illustrate the influence of dissatisfaction and mental activity on dream construction. These findings confirm that the model generates cognitively meaningful and biologically plausible dynamics aligned with known properties of neural networks.

\section{Conclusion}

We present a mathematical model inspired by neurodynamics to explain the process of dream formation, which combines the underlying biological variables through a set of coupled differential equations. The model is designed to simulate various cognitive processes such as emotional arousal, memory suppression, and spontaneous mental activities, and its results are consistent with findings from brain imaging research. The simulation results confirm the validity of the model and demonstrate its potential for use in brain-inspired cognitive artificial systems. In the future, the model may be adapted to real EEG/fMRI data, the system may be extended with random or nonlinear components, and used to investigate different states of consciousness.

\section{References}
P[1], 
author = Marcus E. Raichle, title = The Brain's Default Mode Network, journal = Annual Review of Neuroscience, volume = 38, pages = 433–447, year = 2015
\vspace{0.5cm}

P[2], 
author = Michael C. Anderson, Collin Green, title = Suppressing unwanted memories by executive control, journal = Nature, volume = 410, number = 6826, pages = 366–369, year = 2001
\vspace{0.5cm}

P[3], 
author = Alexander J. Shackman et al., title = The integration of negative affect, pain, and cognitive control in the cingulate cortex, journal = Nature Reviews Neuroscience, volume = 12, pages = 154–167, year = 2011
\vspace{0.5cm}

P[4], 
author = Amit Etkin, Christian Büchel, James J. Gross, title = The neural bases of emotion regulation, journal = Nature Reviews Neuroscience, volume = 12, number = 11, pages = 741–752, year = 2011
\vspace{0.5cm}

P[5], 
author = A. Yelghi and S. Tavangari, title = A Meta-Heuristic Algorithm Based on the Happiness Model, booktitle = Engineering Applications of Modern Metaheuristics, editor = T. Akan and A.M. Anter and A.S¸. Etaner-Uyar and D. Oliva, series = Studies in Computational Intelligence, volume = 1069, year = 2023, publisher = Springer, Cham
\vspace{0.5cm}

P[6], 
author = Kieran C. R. Fox et al., title = The wandering brain: Meta-analysis of functional neuroimaging studies of mind-wandering and related spontaneous thought processes, journal = NeuroImage, volume = 111, pages = 611–621, year = 2015
\vspace{0.5cm}

P[7], 
author = G. William Domhoff, title = The neural substrate for dreaming: Is it a subsystem of the default network?, journal = Consciousness and Cognition, volume = 20, number = 4, pages = 1163–1174, year = 2011
\vspace{0.5cm}

P[8], 
author = Kalina Christoff et al., title = Experience sampling during fMRI reveals default network and executive system contributions to mind wandering, journal = Proceedings of the National Academy of Sciences, volume = 106, number = 21, pages = 8719–8724, year = 2009
\vspace{0.5cm}

P[9], 
author = Mircea Steriade, title = The corticothalamic system in sleep, journal = Frontiers in Bioscience, volume = 6, pages = D137–D148, year = 2001
\vspace{0.5cm}

P[10], 
author = S. Tavangari and Z. Shakarami and A. Yelghi and A. Yelghi, title = Enhancing PAC Learning of Half spaces Through Robust Optimization Techniques, journal = arXiv preprint, year = 2024, volume = arXiv:2410.16573
\vspace{0.5cm}

P[11], 
author = J. Allan Hobson, title = REM sleep and dreaming: towards a theory of protoconsciousness, journal = Nature Reviews Neuroscience, volume = 10, number = 11, pages = 803–813, year = 2009
\vspace{0.5cm}

P[12], 
author = Antti Revonsuo, title = The reinterpretation of dreams: An evolutionary hypothesis of the function of dreaming, journal = Behavioral and Brain Sciences, volume = 23, number = 6, pages = 877–901, year = 2000
\vspace{0.5cm}

P[13], 
author = Aref Yelghi and Shirmohammad Tavangari and Arman Bath,
title = Discovering the characteristic set of metaheuristic algorithm to adapt with ANFIS model, year = 2024
\vspace{0.5cm}

P[14],
author = Larry R. Squire, title = Memory and the hippocampus: A synthesis from findings with rats, monkeys, and humans, journal = Psychological Review, volume = 99, number = 2, pages = 195–231, year = 1992
\vspace{0.5cm}

P[15], 
author = David J. Foster, Matthew A. Wilson, title = Reverse replay of behavioural sequences in hippocampal place cells during the awake state, journal = Nature, volume = 440, number = 7084, pages = 680–683, year = 2006
\vspace{0.5cm}

P[16], 
author = Stanislas Dehaene, Jean-Pierre Changeux, title = Experimental and theoretical approaches to conscious processing, journal = Neuron, volume = 70, number = 2, pages = 200–227, year = 2011
\vspace{0.5cm}

P[17], 
author = Jonathan Smallwood, Jonathan W. Schooler, title = The science of mind wandering: Empirically navigating the stream of consciousness, journal = Annual Review of Psychology, volume = 66, pages = 487–518, year = 2015

\end{document}